\documentclass{article}

\usepackage{graphicx}

\usepackage{hyperref}

\usepackage{amssymb,amsmath}

\usepackage{natbib}

\begin{document}

\title{Type Ia supernova parameter estimation: a  comparison of
two approaches using current datasets}

\author{Bruno L. Lago\thanks{brunolz@if.ufrj.br}\ \thanks{Coordena\c{c}\~ao de Licenciatura em F\'\i sica,  
Centro Federal de
Educa\c{c}\~ao Tecnol\'ogica Celso Suckow da Fonseca, 
CEP 28635-000, Nova
Friburgo, RJ, Brasil}, Maur\'\i cio O. Calv\~ao\thanks{orca@if.ufrj.br}, S\'ergio E. Jor\'as\thanks{joras@if.ufrj.br},\\ Ribamar R. R. Reis\thanks{ribamar@if.ufrj.br}, Ioav Waga\thanks{ioav@if.ufrj.br} and Ram\'on Giostri\thanks{rgiostri@if.ufrj.br}\ \  \thanks{Departamento de
Engenharia Rural, Universidade Federal do Esp\'\i rito
Santo, C. P. 16, CEP 29500-000, Alegre, ES, Brasil}\\
\textit{Instituto de F\'\i sica, Universidade
Federal do Rio de Janeiro,}\\
\textit{C. P. 68528, CEP 21941-972, Rio de Janeiro, RJ,
Brasil}}

\date{}

\maketitle

\begin{abstract}

By using the Sloan Digital Sky Survey (SDSS) first year type Ia
supernova (SN Ia) compilation, we compare two different approaches 
(traditional $\chi^2$ and complete likelihood) to determine parameter 
constraints when the magnitude dispersion is to be estimated as well. We
consider
cosmological constant + Cold Dark Matter ($\Lambda$CDM) and spatially flat,
constant $w$ Dark Energy + Cold Dark Matter (F$w$CDM) cosmological models and
show that, for current data, there is a small difference in the best fit values
and $\sim$ 30\% difference in confidence contour areas in case the MLCS2k2
light-curve fitter is adopted. For the SALT2 light-curve fitter the differences
are less significant ($\lesssim 13\%$ difference in areas). In both cases
the likelihood approach gives more restrictive constraints.  We argue for the
importance of using the complete likelihood instead of the $\chi^2$ approach
when dealing with parameters in the expression for the variance.

\end{abstract}

\section{Introduction}
\label{sec:introduction}

By the end of the last century observations of type Ia supernovae (SNe 
Ia), used as standard candles, directly established the
acceleration of the universe
\citep{Riess1998,Perlmutter1999}, an
awesome result, possibly only surpassed by the discovery of its very own
expansion, around eighty years ago. They are still the
backbone for the
prevailing $\Lambda$CDM concordance model, which is  furthermore corroborated by
the combination of
other probes, such as, e.g., cosmic microwave background anisotropies baryon
acoustic oscillations and 
galaxy clustering \citep{Komatsu2011,Eisenstein2005,Percival2010,Vikhlinin2009,Mantz2010}.

The sample standard deviation in the inferred, uncorrected absolute $B$
magnitude of typical sets 
of SNe Ia is of the order 0.4 mag (even after the exclusion of
outliers such as the overluminous 1991T-like SNe Ia and the underluminous
1991bg-like SNe Ia) \citep{Phillips1993,Vaughan1995,Hamuy1996,Contardo2000}.  This
amounts to a fractional uncertainty in luminosity of $-30\%$ and
(under the assumption of negligible uncertainties in all other
quantities) in luminosity distance of $20\%$. It is thus clear that SNe Ia are
not exactly prima facie standard candles at all. 
However, even such a scatter, as compared to other categories of astrophysical
sources (other supernova types, gamma-ray bursts, etc), is small and, in
conjunction with their typical high peak luminosity 
($10^{36}$ W$\sim 4\times 10^9L_\odot$), justifies the effort to improve their
use as cosmological
\emph{standardizable} candles. After some
phenomenological standardization recipes
(cf. Section \ref{sec:light_curve_fitting}), the scatter in $M_B$ decreases to
around 0.15 mag, which amounts to a fractional uncertainty in luminosity of
$-13\%$
and in luminosity distance of $7\%$.

Already in the landmark papers from the two original surveying groups
\citep{Riess1998,Perlmutter1999}, a
legitimate concern was expressed about the presence of not properly
accounted for systematic effects. At that time, however, 
there was only a small number (the order of tens) of observed SNe Ia in any of
the
available samples, so the uncertainties were statistically limited; the main
task was gathering new, larger, uniform datasets. For the recent compilations 
and surveys
\citep{Wood-Vasey2007,Kowalski2008,Hicken2009,Conley2011,Kessler2009,Amanullah2010}, and the more so for the future
ones
\citep{Pan-STARRS, DES, LSST},
we are no longer sample-limited; the urgency is again on the physics of the
SN Ia phenomenon and all the aspects which impact it: nature of progenitor
binary system (single-degenerate versus double degenerate, tardy versus prompt
events) 
\citep{Dilday2010,Hayden2010,Maoz2010},
properties of host galaxy \citep{Sullivan2010,Lampeitl2010,Gupta2011},
mechanisms of explosion \citep{Hillebrandt2000}, extinction/intrinsic color
variations 
\citep{Nobili2009,Yasuda2010},
K-correction and template calibration \citep{Nugent2002,Hsiao2007}, flux calibration
\citep{Faccioli2011},
inhomogeneities (peculiar velocities
\citep{Hui2006,Davis2010}, 
gravitational lensing
\citep{Wang2005b,Kronborg2010}, etc). We would
also like to call attention to the particularly clear, informative and
up-to-date generic review articles on SNe Ia by \citet{Kirshner2010}, \citet{Howell2011},
and \citet{Goobar2011}.
Simultaneously with this physical endeavor from first principles, we should also
exercise our best
statistical consistent practices to analyze and mine the data and to test the
robustness of our inferences. It is to the latter that our paper is devoted.

In contrast to the traditional $\chi^2$ approach present in most of the
literature,
in this work we discuss a different approach to parameter fitting
based on the likelihood. We call attention to the fact that, when we want to
estimate both the covariance and the mean of a Gaussian process, the
ordinary (uncorrected) $\chi^2$ approach (or any iterative recipe
therefrom, for that matter) cannot be straightforwardly applied, lest we
might lose a nontrivial term in the objective function to be
extremized. This is necessary when the covariance itself
does depend on free parameters of the underlying model. 
Recently, similar criticisms to the traditional $\chi^2$ method were presented
by Kim \citep{Kim2011}  (see also \citep{Vishwakarma2010}). 
In \citep{Kim2011} a likelihood approach to simulated data for the distance
moduli was used, and the focus was on the determination
 of the intrinsic dispersion,  without reference to a particular light-curve
fitter. In our work, by using 
real-data, the problem is reconsidered in the context 
of the MLCS2K2 and SALT2 light-curve fitters. In particular, we point out that,
in principle, with
SALT2 additional difficulties could arise due to the astrophysical parameters. 
We compare the results of  the traditional $\chi^2$ treatment to the likelihood
one. We restrict 
ourselves to  SNe Ia datasets only, without taking into account other important
sources of 
information (such as CMB, BAOs or clusters) in order to make our point pristine,
by avoiding the
masking due to any other such probes.

The structure of the paper is as follows. In Section
\ref{sec:light_curve_fitting}, 
we briefly remind the general procedure to go from the raw data to the final
estimated parameters. In Section \ref{sec:least_squares}, we describe the two
most widely used ``fitter pipelines'' (MLCS2k2 and SALT2), which use the
traditional $\chi^2$ approach. In Section \ref{sec:likelihood}, we critically
review the
aforementioned traditional approach and present the likelihood-based one. In
Section \ref{sec:results}, the main numerical results are shown, for
both fitters, in the case of the Sloan Digital Sky Survey (SDSS) first year
compilation \citep{Kessler2009}. Finally, in Section \ref{sec:conclusions}, we
end up with some
discussions and conclusions.

\section{Light-curve fitting}
\label{sec:light_curve_fitting}

The ``primary'' data of any SN Ia survey are apparent magnitudes (or fluxes)
in a given set of filters, at a series of epochs (phases), for each supernova.
This constitutes an array $m_{i,Y,\gamma}$ ($f_{i,Y,\gamma}$), where $i$ labels
the supernova, $Y$ labels the
filter (or band) and $\gamma$ labels the epoch
in the time series. For a given supernova $i$, observed in a given filter $Y$,
the scatterplot of the points 
$(t_{\gamma},m_{i,Y,\gamma})$, where $t_\gamma$ is the observed time, is what we
will call (a sampling of) the raw light curve.

As mentioned in Section \ref{sec:introduction},
SNe Ia are \emph{standardizable} candles; this means there is a 
phenomenological recipe whereby the raw light curves, after being
subjected to a transformation by a $N_{\rm st}$-parameter function, furnish a
new set of so-called \emph{standardized} light curves; this means the
dispersion in the new
magnitudes is considerably smaller than in the original
input set. The aforementioned phenomenological recipe is not unique at all:
there are several light-curve fitters in the literature
\citep{Jha2007,Wang2003,Wang2005a,Guy2005,Guy2007,
Conley2008,Burns2011}. Here we will exploit some features
of the two most common ones: the Multicolor Light-Curve Shape (MLCS2k2)
\citep{Jha2007} one and the Spectral
Adaptive Light curve Template (SALT2) \citep{Guy2007}.

In a nutshell:
\begin{itemize}
\item The MLCS2k2 fitting model \citep{Jha2007} describes the variation among SNe
Ia light
curves with a single parameter ($\Delta$). Excess color variations
relative to the one-parameter model are assumed
to be the result of extinction by dust in the host galaxy
and in the Milky Way. The MLCS2k2 theoretical magnitude, observed in
an arbitrary filter $Y$, at an epoch $\gamma$,
is given by
\begin{equation}
m_{Y,\gamma}^{\rm th}=M_{Y',\gamma}+p_{Y',\gamma}\Delta
+q_{Y',\gamma}\Delta^2
+K_{Y'Y,\gamma}+\mu +X_{Y',\gamma}^{\rm host} +X_{Y,\gamma}^{\rm MW},
\label{mlcs2k2model}
\end{equation}
where $Y' \in \{U, B, V, R, I\}$ is one of the supernova rest-frame filters for 
which the model is defined, $\Delta$ is the MLCS2k2 shape-luminosity 
parameter that accounts for the correlation between peak luminosity
and the shape/duration of the light curve. Furthermore, the model for 
the host-galaxy extinction is 
$X^{\rm host}_{Y',\gamma}=\zeta_{Y',\gamma}(a_{Y'}+b_{Y'}/R_V)A_V$, 
where $\zeta_{Y',\gamma}:=X^{\rm host}_{Y',\gamma}/X^{\rm host}_{Y',0}$, 
and $a_{Y'}$, $b_{Y'}$ are constants; as usual, $A_V$ is the $V$ band
extinction, at $B$ band peak 
($a_V=1$, $b_V=0$), and $R_V:=A_V/E(B-V)$, the ratio of $V$ band extinction
to color excess, at $B$ band peak. Finally, $X^{\rm MW}_{Y,\gamma}$ is the
Milky Way extinction, $K_{Y'Y,\gamma}$ is the $K$-correction between rest-frame
and
observer-frame filters, and $\mu$ is the distance modulus.
The coefficients $M_{Y',\gamma}$, $p_{Y',\gamma}$, and $q_{Y',\gamma}$
are model vectors that have been evaluated using nearly 100 well observed
low-redshift SNe Ia as a training set. $\gamma=0$ labels quantities at the
$B$ band peak magnitude epoch.

Fitting the model to each SN Ia magnitudes, usually fixing $R_V$ 
gives $\mu$, $\Delta$, $A_V$ and $t_0$, the $B$-band peak magnitude
epoch.

\item The SALT2 fitter \citep{Guy2007} makes use of a
two-dimensional
surface in time and wavelength that describes the temporal evolution
of the de-redshifted (rest-frame) spectral energy distribution (SED) or specific
flux (flux per unit wavelength) for SNe Ia. The
original model was trained on a set of combined light-curves and 303 spectra,
not only from (very) nearby but also medium and high redshift SN Ia.

In SALT2, the de-redshifted (rest-frame) specific flux at
wavelength $\lambda$ and phase (time) $p$ ($p = 0$ at $B$-band maximum) is
modeled by
\begin{equation}
\phi(p,\lambda; x_0,x_1,c)=x_0[M_0(t,\lambda)+x_1M_1(t,\lambda)]
\exp [cC(\lambda)].
\label{salt2restflux}
\end{equation}
and does depend, through the parameters $x_0, x_1$, and $c$ on the
particular type Ia supernova. $M_0(t, \lambda)$, $M_1(t, \lambda)$, and
$C(\lambda)$ are determined from the
training process described in \citep{Guy2007}. $M_0(t, \lambda)$, $M_1(t,
\lambda)$
are the zeroth and the first moments of the distribution of 
training sample SEDs. One might consider adding
moments of higher order to Eq. (\ref{salt2restflux}).

To compare with photometric SNe Ia data, the
(unredshifted) observer-frame
flux in passband $Y$ is calculated as
\begin{equation}
F^Y(p(1+z))=(1+z)\int d\lambda'\left[\lambda'
\phi(p,\lambda')T^Y(\lambda' (1+z))\right],
\label{salt2obsflux}
\end{equation}
where $T^Y(\lambda)$ defines the transmission curve of the observer-frame
filter $Y$, and $z$ is the redshift.

As called attention to above, each SN Ia
light curve is fitted separately using Eqs. (\ref{salt2restflux}) 
and (\ref{salt2obsflux}) to determine
the parameters $x_0$, $x_1$, and $c$ with corresponding errors. However, the
SALT2 
light-curve fit does not yield an independent distance
modulus estimate for each SN Ia. As we will see in the next section, the
distance moduli are estimated as part of a global parameter fit
to
an ensemble of SN Ia light curves in which cosmological parameters
and global SN Ia properties are also obtained.
\end{itemize}

In the next section we will discuss how to obtain constraints on cosmological
parameters
using MLCS2k2 and SALT2 output quantities as our data.

\section{The traditional $\chi^2$ approach}
\label{sec:least_squares}

The prevailing SNe Ia cosmological analysis is based on the $\chi^2$ function:
\begin{equation}
\label{chi2_base}
\chi ^2:=\boldsymbol{X}^T \boldsymbol{\Sigma}^{-1}\boldsymbol{X}
\end{equation}
where $\boldsymbol{X}:=(\boldsymbol\mu-\boldsymbol{\mu}_{th}(\boldsymbol{z},
\boldsymbol{\Theta}))$, $\boldsymbol{\mu}$ is the set of
distance moduli derived from the light curve fitting procedure for each SN Ia
event, 
at redshifts given by $\boldsymbol{z}$, $\boldsymbol{\mu}_{th}(\boldsymbol{z},
\boldsymbol{\Theta})$ is the theoretical prediction for them, given in terms of 
a vector $\boldsymbol{\Theta}$ of parameters and
$\boldsymbol{\Sigma}$ is the covariance matrix of the events.

As discussed in the previous section, each light curve fitter 
gives a different set of output or processed data, which are 
not related to the distance modulus in the same way. In this
work, to construct the traditional $\chi^2$ function of this section or the
proposed likelihood function of the next section,
we consider as ``data'' the distance modulus estimation obtained from 
MLCS2k2 and SALT2 processing.

\subsection{The $\chi^2$ approach from SALT2 output}

The SALT2 light curve fitter gives three quantities, with
corresponding errors, 
to be used in the analysis of cosmology:\\
\begin{equation}
 m^{\ast}_B:=-2.5\log \left[x_0\int d\lambda'
M_0(p=0,\lambda')T^B(\lambda')\right]\,,
\end{equation}
to be interpreted as the peak de-redshifted (rest-frame) magnitude in the $B$
band, $x_1$, a parameter related to the stretch of the light-curve and
$c$, related to the color of the supernova alongside the redshift $z$ of the
supernova. The distance modulus is modelled, in this context, as
a function of ($m_B^{\ast}$, $x_1$, $c$) and
two new parameters, $\boldsymbol{\delta}:=(\alpha,\beta)$, plus the peak
absolute magnitude,
in $B$ band, $M_B$.
Defining the corrected magnitude as
\begin{equation}
m^\mathrm{corr}_B(\boldsymbol{\delta}):=m_B^{\ast}+\alpha x_1 -\beta c
\label{m_corr}
\end{equation}
we can write
\begin{equation}
\mu(\boldsymbol{\delta},M_B) = m^\mathrm{corr}_B(\boldsymbol{\delta}) - M_B
\end{equation}

Assuming that all SNe Ia events are 
independent, one can rewrite Eq. (\ref{chi2_base}) as 
\begin{equation}
\chi^2_{\mathrm{SALT2}}(\boldsymbol{\theta},
\boldsymbol{\delta}, {\cal M}(M_B, h)) = \sum_{i=1}^N
\frac{[\mu_{i}(\boldsymbol{\delta},M_B)-\mu_\mathrm{th}(z_i;
\boldsymbol{\theta}, h)]^2}{\sigma_i^2(\boldsymbol{\delta}) +
\sigma_\mathrm{int}^ 2}
\label{chi2_salt2}
\end{equation}
where $N$ is the number of SNe Ia in the sample, $\boldsymbol{\theta}$ denotes
the cosmological parameters other than $h$, with the present value of the Hubble
parameter given by 
$H_0= 100 h \mbox{ km}\cdot\mbox{s}^{-1}\cdot\mbox{Mpc}^{-1}$.
The theoretical distance modulus is given by
\begin{equation}
\mu_\mathrm{th}(z; \boldsymbol{\theta},
h)=5\log[\mathcal{D}_L(z; \boldsymbol{\theta})]+\mu_0(h)\,.
\label{m_th_def}
\end{equation}
with
\begin{equation}
\mu_0(h) := 5\log\left( \frac{100c/(\mbox{km/s})}{h} \right)\,.
\end{equation}
The dimensionless luminosity distance (in units of the Hubble distance
today), $\mathcal{D}_L$, for comoving observers in a
Robertson-Walker universe, is given by
\begin{equation}
\mathcal{D}_L(z; \boldsymbol{\theta})  = \left\{\begin{array}{ll}
 (1+z)\left(\frac{1}{\sqrt{\Omega_{k0}}}\right)\sinh\left(\sqrt{\Omega_{k0}}{
\int_{z'=0}^z\frac{1}{E(z'; \boldsymbol{\theta})}dz'}\right), &\hbox{if}\;
\Omega_{k0}>0,  \\
 (1+z){\int_{z'=0}^z\frac{1}{E(z'; \boldsymbol{\theta})}dz'}, &\hbox{if}\;
\Omega_{k0}=0,  \\
 (1+z)\left(\frac{1}{\sqrt{-\Omega_{k0}}}\right)\sin\left(\sqrt{-\Omega_{k0}}{
\int_{z'=0}^z\frac{1}{E(z'; \boldsymbol{\theta})}dz'}\right), &\hbox{if}\;
\Omega_{k0}<0,
 \end{array} \right.
\end{equation}
where $\Omega_{k0}$ is the ``curvature density parameter''
(whatever the underlying dynamical gravitational theory), such that it is
proportional to the three-curvature, and 
\begin{equation}
 E(z;
\boldsymbol{\theta}):=H(z;\boldsymbol{\theta},h)/H_0
\end{equation}
is the dimensionless Hubble parameter. As indicated in Eq. (\ref{chi2_salt2}),
the $\chi^2$ function depends on the parameters $M_B$ and $h$ only through
their combination
\begin{equation}
 {\cal M}(M_B, h) := M_B + \mu_0(h)\,.
\end{equation}
It may thus be thought of as effectively directly dependent on
only the parameters $\boldsymbol{\theta}$, $\boldsymbol{\delta}$ and ${\cal
M}$.

A floating dispersion term, $\sigma_{int}$, ``which contains potential
sample-dependent systematic errors that have not
been accounted for and the observed intrinsic SNe Ia dispersion''
\citep{Amanullah2010},
is added in quadrature to the distance modulus dispersion, which is given
by
\begin{equation}
\sigma_i^2(\boldsymbol{\delta})=\sigma^
2_{m_B^*,i}+\alpha^2\sigma_{x_1,i}^2 + \beta^2\sigma_{c,i}^2 
+2\alpha\sigma_{m_B^*,x_1,i}-2\beta\sigma_{
m_B^*,c,i} -2\alpha\beta\sigma_{x_1,c,i} +\sigma^2_{\mu,z,i}.
\label{sigsalt2}
\end{equation}
where $\sigma^2_{\mu,z,i}$ is the contribution to the distance modulus
dispersion due to redshift uncertainties from peculiar velocities
and also from the measurement itself. Following \citep{Kessler2009}, 
we will model this contribution, for simplicity, using the distance-redshift 
relation for an empty universe which gives
\begin{equation}
\sigma_{\mu,z,i}=\sigma_{z,i}\left(\frac{5}{\ln
10}\right)\frac{1+z_i}{z_i(1+z_i/2)},
\end{equation}
with
\begin{equation}
\sigma_{z,i}^2=\sigma_{spec,i}^2+\sigma_{pec}^2,
\end{equation}
where $\sigma_{spec,i}$ is the redshift measurement error,
and  $\sigma_{pec}=0.0012$ is the redshift uncertainty due to peculiar
velocity.

As advocated by some groups \citep{Astier2006}, minimizing Eq.
(\ref{chi2_salt2}) 
gives rise to a bias towards increasing values of $\alpha$ and $\beta$. 
In order to circumvent this feature an iterative method is performed,
according to their approach.

In this iterative method, the $\chi^2$ presented in Eq. (\ref{chi2_salt2}) is
replaced by
\begin{equation}
\chi^2_{\mathrm{SALT2}}(\boldsymbol{\theta}, \boldsymbol{\delta}, {\cal M})
= \sum_{i=1}^N \frac{[\mu_{i}(\boldsymbol{\delta},
M_B)-\mu_\mathrm{th}(z_i;\boldsymbol{\theta},h)]^2}{\sigma_i^2(\boldsymbol{\eta}
) + \sigma_\mathrm{int}^ 2}\,.  
\label{chi2_salt2_new}
\end{equation}
Notice that, in this expression, $\boldsymbol{\eta}$ is not a parameter of the
$\chi^2_{\mathrm{SALT2}}$. In order to obtain the best fit values for the
parameters,
$\boldsymbol{\eta}$ is given initial values and the optimization is performed on
$\boldsymbol{\theta}, \boldsymbol{\delta}$ and ${\cal M}$. After this step,
$\boldsymbol{\eta}$ is updated 
with the best fit value of $\boldsymbol{\delta}$ and the optimization is
performed again. 
The process continues until a convergence is achieved, which means that
$\boldsymbol{\eta}$ 
does not change under the required precision.

During this process $\sigma_\mathrm{int}$ is not
considered as a free parameter to be optimized, being determined rather by the
following procedure: start with a guess value
(usually $\sigma_\mathrm{int}=0.15$). Perform the iterative procedure described
above. The value of $\sigma_\mathrm{int}$ is then obtained by fine tuning it so
that the reduced $\chi^2$ equals unity (with all the other parameters fixed on
their best fit values). The iterative procedure is repeated once more with this
new value and the final best fit values are obtained.  It is important to note 
that the value of $\sigma_{\rm int}$
affects both the best fit and the confidence levels of the parameters,
since it changes the weight given to each supernova in the $\chi^2$ [cf. Eq. 
(\ref{chi2_salt2_new})].

\subsection{The $\chi^2$ approach from MLCS2k2 output}

The MLCS2k2 light curve fitter is also a distance estimator and gives us
directly a cosmology-independent estimation of the distance modulus.
In this context, the analogue of Eq. (\ref{chi2_salt2}) is
\begin{equation}
\chi^2_{\mathrm{MLCS2k2}}(\boldsymbol{\theta},h) = \sum_i^N
\frac{[\mu_{i}-\mu_\mathrm{th}(z_i;\boldsymbol{\theta},h)]^2}{\sigma_i^2 +
\sigma_\mathrm{int}^ 2+\sigma^2_{\mu,z,i}}
\label{chi2_mlcs2k2}
\end{equation}
where $\sigma_i$ is the distance modulus dispersion as given by MLCS2k2.

The procedure to obtain $\sigma_\mathrm{int}$ is similar to the one described in
the previous Subsection, however, in this case we use only a subsample of
nearby 
SNe Ia and not the full one, as for the SALT2 analysis. After setting up the 
value of $\sigma_\mathrm{int}$, we minimize the $\chi^2_{\mathrm{MLCS2k2}}$
using the full SNe Ia
sample to obtain the best fit values for $\boldsymbol{\theta}$ and $h$.

\section{The proposed likelihood approach}
\label{sec:likelihood}

Considering the SNe Ia light curve fitting parameters as Gaussian distributed
random variables, we propose to take as starting point the likelihood
\begin{equation}
L=\frac{1}{\sqrt{(2\pi)^N\hbox{det}\,\boldsymbol{\Sigma}}}\exp(-\boldsymbol{X}^T
\boldsymbol{\Sigma}^{-1}\boldsymbol{X}/2)\,,
\label{like}
\end{equation}
which is related to the $\chi^2$ in Eq. (\ref{chi2_base}) by
\begin{equation}
\mathcal{L}:=-2\ln L=\chi^2  + \ln
\hbox{det}\,\boldsymbol{\Sigma}+ N\ln (2\pi)\,.
\label{loglike}
\end{equation}
Eqs. (\ref{like}) and (\ref{loglike}) are the single basis upon 
which our whole statistical procedure lies. 
When the full covariance of the problem is known, minimizing $\chi^2$ is
completely
equivalent to minimizing $\mathcal{L}$. However, this is not the case for
current SNe Ia 
observations and neglecting the last but one term in Eq. (\ref{loglike}) would,
in principle, 
lead to a biased result. Our proposal is minimizing the following functions
for each case discussed in the previous section
\begin{equation}
\label{like_salt}
\mathcal{L}_{\mathrm{SALT2}}(\boldsymbol{\theta}, \boldsymbol{\delta}, {\cal M},
\sigma_\mathrm{int})=\chi^2_{\mathrm{SALT2}}(\boldsymbol{\theta},
\boldsymbol{\delta}, {\cal M}, \sigma_\mathrm{int})
+\sum_i^N\ln(\sigma_i^2(\boldsymbol{\delta}) + \sigma_\mathrm{int}^ 2)
\end{equation}
and
\begin{equation}
\label{like_mlcs}
\mathcal{L}_{\mathrm{MLCS2k2}}(\boldsymbol{\theta}, h,
\sigma_\mathrm{int})=\chi^2_{\mathrm{MLCS2k2}}(\boldsymbol{\theta}, h,
\sigma_\mathrm{int})
+\sum_i^N\ln(\sigma_i^2 +\sigma_\mathrm{int}^ 2),
\end{equation}
where we neglected parameter-independent terms. 
$\chi^2_{\mathrm{SALT2}}(\boldsymbol{\theta}, \boldsymbol{\delta}, {\cal M},
\sigma_\mathrm{int})$ and
$\chi^2_{\mathrm{MLCS2k2}}(\boldsymbol{\theta}, h, \sigma_\mathrm{int})$ are
given, respectively,
by Eqs. (\ref{chi2_salt2}) and (\ref{chi2_mlcs2k2}) now considering
$\sigma_\mathrm{int}$ \emph{also as a free parameter}. With this procedure, we
can
obtain directly unbiased probability distributions functions for all parameters,
including 
$\sigma_\mathrm{int}$ and $\boldsymbol{\delta}$.

\section{Results}
\label{sec:results}

In this section we compare the results obtained from the $\chi^2$
and the likelihood approaches, as described in Sections
\ref{sec:least_squares} and \ref{sec:likelihood}, using real data
from the SDSS first year compilation \citep{Kessler2009}.

In order to perform the comparison, we considered the following cosmological
models:
\begin{itemize}
\item $\Lambda$CDM, in which we can write the Friedmann equation, in terms of
the parameters $\boldsymbol{\theta}=(\Omega_{m0},\Omega_{k0})$, as
\begin{equation}
E^2(z; \boldsymbol{\theta})= \Omega_{m0}(1+z)^3 +\Omega_{k0}(1+z)^2
+(1-\Omega_{m0}-\Omega_{k0}).
\end{equation}

\item F$w$CDM, described by $\boldsymbol{\theta}=(\Omega_{m0},w)$ and
\begin{equation}
E^2(z; \boldsymbol{\theta})= \Omega_{m0}(1+z)^3 +
%  \Omega_{k0}(1+z)^2 +
(1-\Omega_{m0}-\Omega_{k0})(1+z)^{3(1+w)}.
\end{equation}

\end{itemize}

We chose the $\Lambda$CDM and F$w$CDM models to directly compare the best-fit
and the
68\% and 95\% confidence contours for the parameters $\boldsymbol{\delta}$, 
$\sigma_{\text{int}}$ and $\boldsymbol{\theta}$, for both SALT2 and  
MLCS2k2 data.
The best fit values were obtained with the MIGRAD minimization of the
Minuit \citep{James1975} implementation in ROOT \citep{Antcheva2009} and
the probability distributions were obtained
with Monte Carlo Markov Chains (MCMC). We considered as the probability 
distributions, in the context of $\chi^2$ approach, the following functions:
\begin{eqnarray}
\mathcal{P}_{SALT2}(\boldsymbol{\theta},\boldsymbol{\delta},\mathcal{M})
& = &
N_{SALT2}\exp\left[-\chi^2_{SALT2}(\boldsymbol{\theta},\boldsymbol{\delta},
\mathcal{M},
\sigma_{\text{int}})/2\right],  \\
\mathcal{P}_{MLCS2k2}(\boldsymbol{\theta},h)
& = & N_{MLCS2k2}\exp\left[-\chi^2_{MLCS2k2}(\boldsymbol{\theta},h,
\sigma_{\text{int}})/2\right],
\end{eqnarray}
where the normalization factors $N_{SALT2}$ and $N_{MLCS2k2}$ are independent
of the parameters to be estimated. Note that, for the traditional $\chi^2$
method, $\sigma_{\text{int}}$ 
is fixed so the probability
distribution does not depend on it.

\begin{figure}
\begin{centering}
\includegraphics[width=0.49\textwidth]{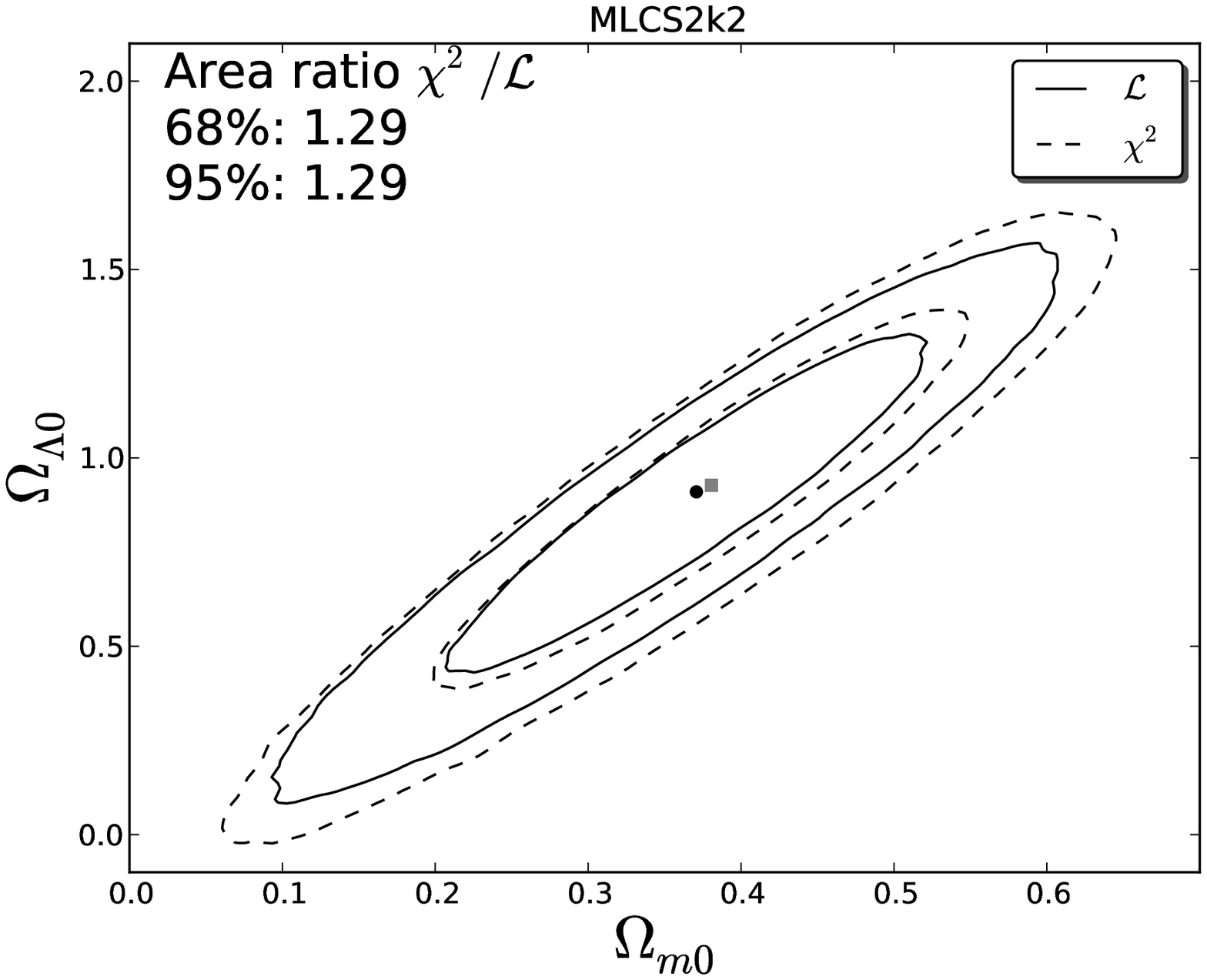}
\includegraphics[width=0.49\textwidth]{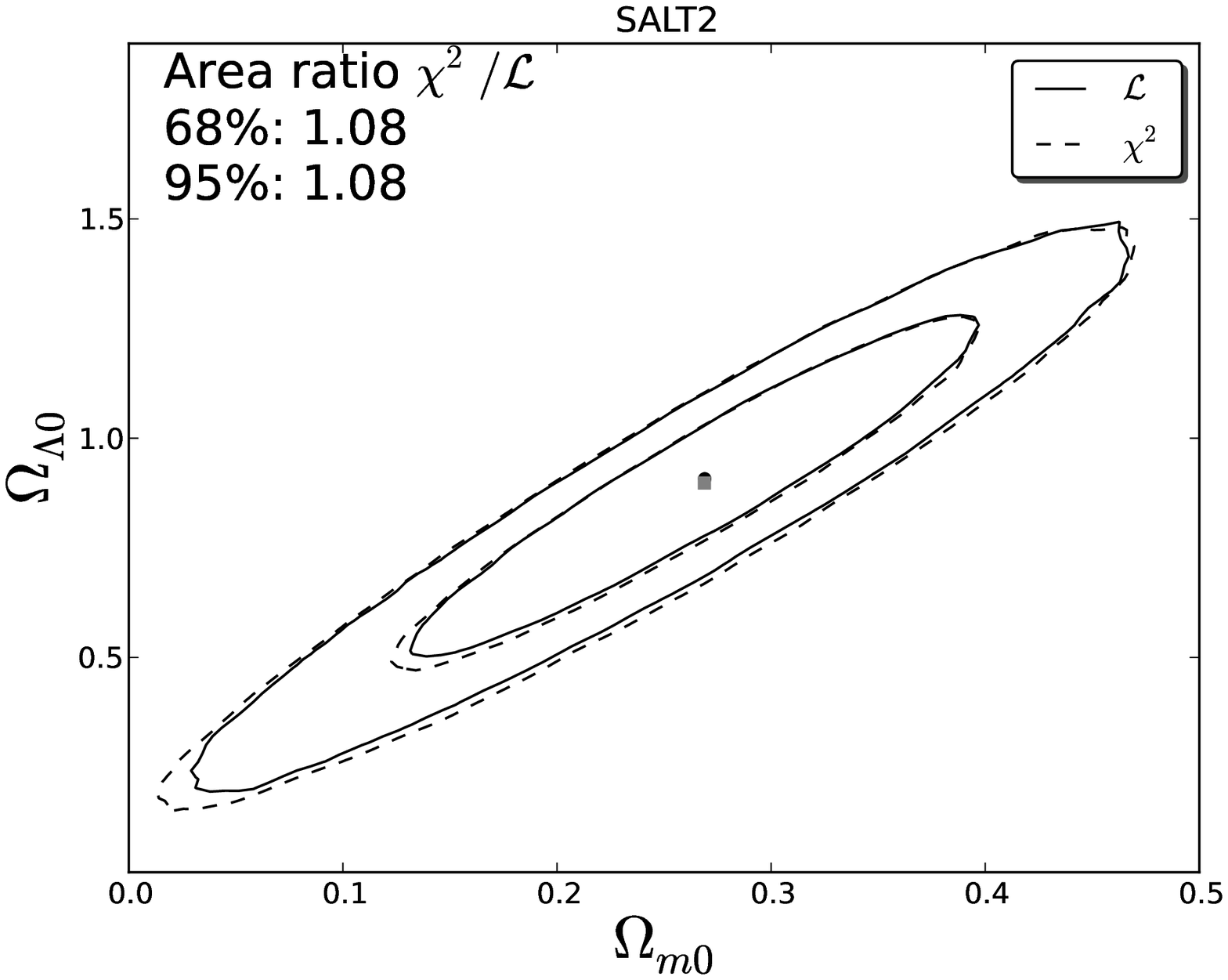}
\par\end{centering}

\caption{68\% and 95\% confidence level contours, in the plane
$\Omega_{m0}\times\Omega_{\Lambda0}$, for the $\Lambda$CDM
model. We marginalized over all other parameters with flat prior. 
The solid (dashed) lines are the results for the likelihood
($\chi^2$) approach. The black circle (gray square) is the best fit value
for the likelihood ($\chi^2$).
\textit{Left panel:} Results for the MLCS2k2 version of the SDSS compilation.
\textit{Right panel:} Results for the SALT2 version of the SDSS compilation.  
\label{fig:SDSS_LCDM_cosmopar}}

\end{figure}

\begin{figure}
\begin{centering}
\includegraphics[width=0.49\textwidth]{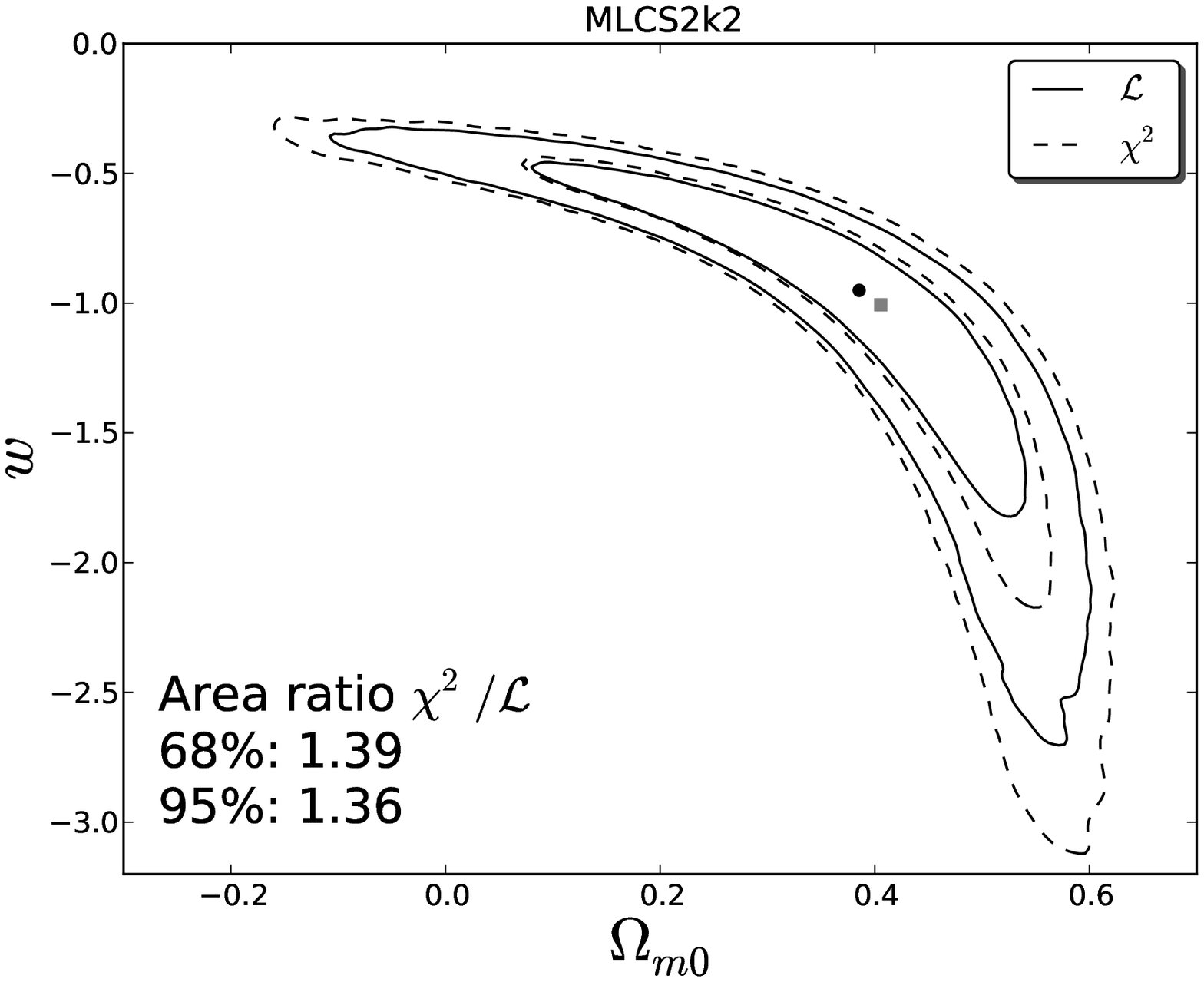}
\includegraphics[width=0.49\textwidth]{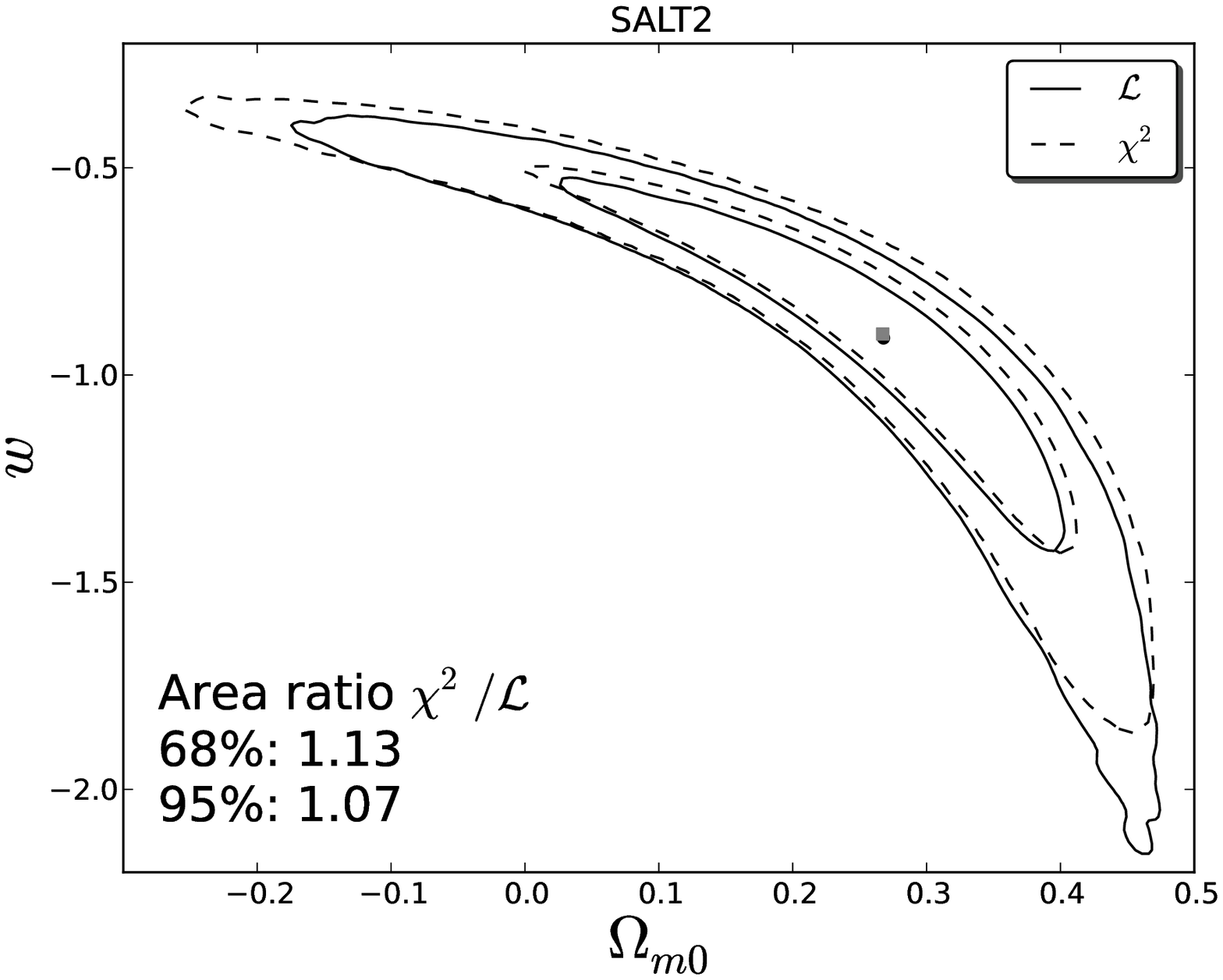}
\par\end{centering}

\caption{68\% and 95\% confidence level contours, in the plane
$\Omega_{m0}\times w$, for the F$w$CDM
model. We marginalized over all other parameters with flat prior. 
The solid (dashed) lines are the results for the likelihood
($\chi^2$) approach. The black circle (gray square) is the best fit value
for the likelihood ($\chi^2$).
\textit{Left panel:} Results for the MLCS2k2 version of the SDSS compilation.
\textit{Right panel:} Results for the SALT2 version of the SDSS compilation.  
\label{fig:SDSS_wCDM_cosmopar}}

\end{figure}

\begin{figure}
\begin{centering}
\includegraphics[width=0.49\textwidth]{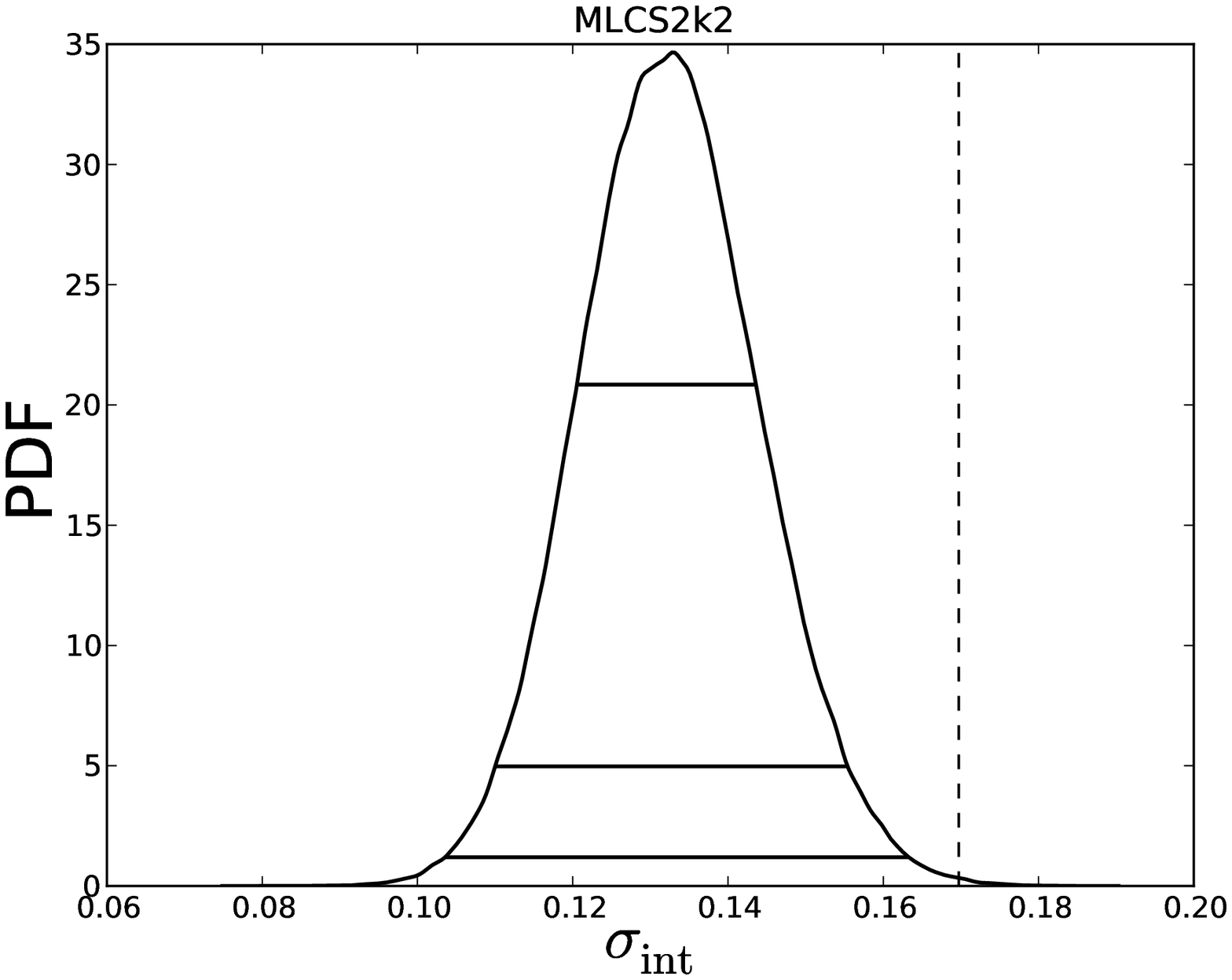}
\includegraphics[width=0.49\textwidth]{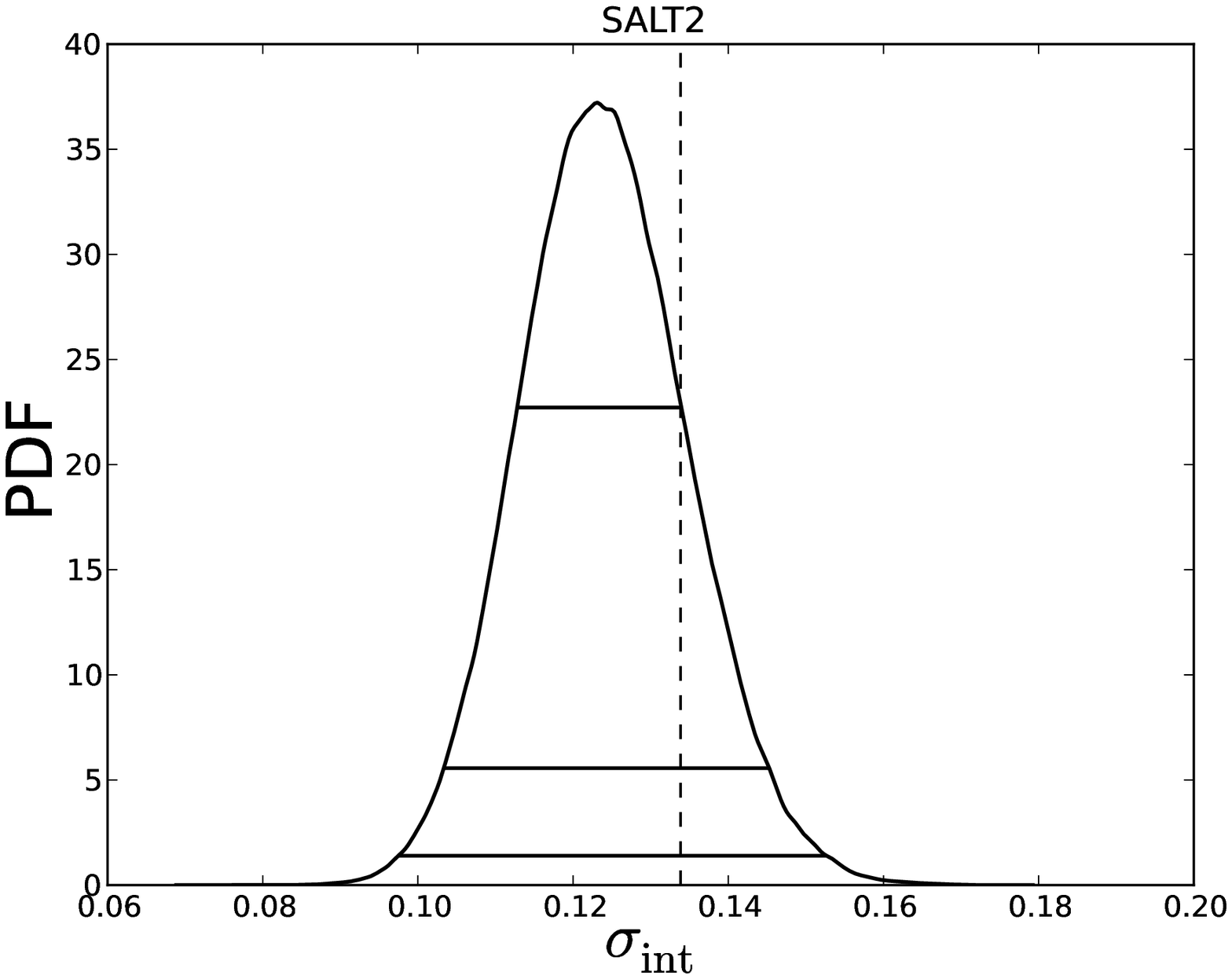}
\par\end{centering}

\caption{Distributions for $\sigma_{\text{int}}$ for the
$\Lambda$CDM model from the likelihood approach. 
We marginalized over all other parameters with flat prior.
The horizontal lines depict the 68\%, 95\% and 99\% confidence
intervals. 
The dashed line is the result for the $\chi^2$ approach.
\textit{Left panel:} Results for the MLCS2k2 version of the SDSS compilation.
\textit{Right panel:} Results for the SALT2 version of the SDSS compilation.  
\label{fig:SDSS_sint_dist}}

\end{figure}

In Figs. \ref{fig:SDSS_LCDM_cosmopar} and \ref{fig:SDSS_wCDM_cosmopar}
we show the confidence contours for the parameters $\boldsymbol{\theta}$ for
 $\Lambda$CDM and F$w$CDM models, respectively. For these models, we note that 
the differences between the best fit and the area of the contours, for the 
$\chi^2$ 
and likelihood approaches, are more significant when the MLCS2k2 fitter is
used. 
In fact, for the SALT2 fitter, the differences are not significant (less
than $13\%$) --- see also Fig.~\ref{fig:SDSS_sint_dist} and discussion below. If
this is a general feature or depends on the models or dataset used has to be
further investigated.

\begin{figure}
\begin{centering}
\includegraphics[width=0.49\textwidth]{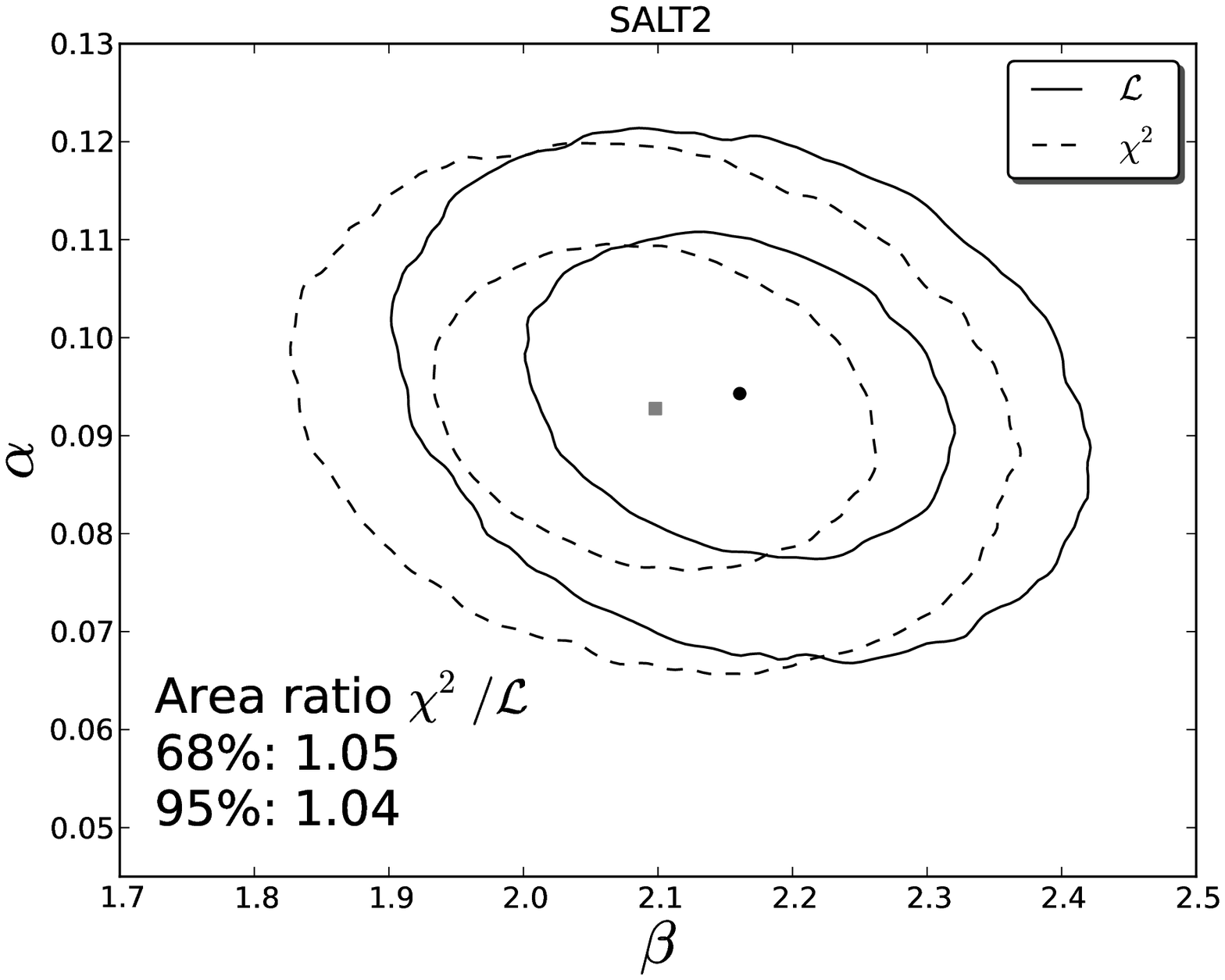}
\includegraphics[width=0.49\textwidth]{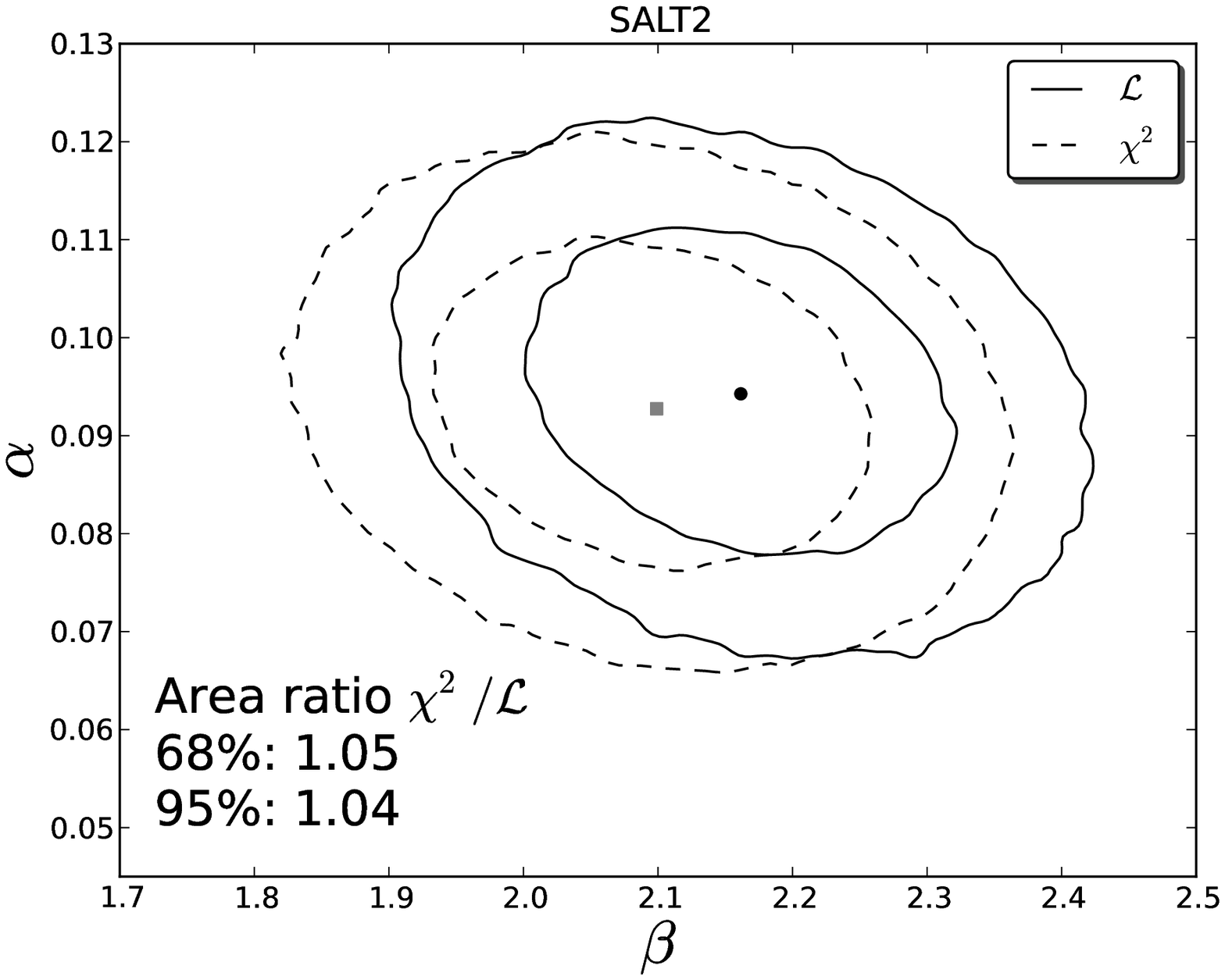}
\par\end{centering}

\caption{68\% and 95\% confidence level contours in the plane
$\beta\times\alpha$. We marginalized over all other parameters with flat prior. 
The solid (dashed) lines are the results for the likelihood
($\chi^2$) approach. The black circle (gray square) is the best fit value
for the likelihood ($\chi^2$).
\textit{Left panel:} Results for the $\Lambda$CDM model with the SDSS
compilation.
\textit{Right panel:} Results for the F$w$CDM model with the SDSS compilation.  
\label{fig:SDSS_beta_alpha}}

\end{figure}

In Fig. \ref{fig:SDSS_beta_alpha} we show the confidence contours for
the SALT2 parameters $\boldsymbol{\delta}$ for both $\Lambda$CDM and
F$w$CDM models. We can see that there is no significant difference
in the constraints for $\alpha$. For $\beta$ we find a  bias that is, however, 
small compared to the 68\% confidence interval for this parameter.

In Fig. \ref{fig:SDSS_sint_dist} we show the probability distributions 
for $\sigma_{\text{int}}$,
given by the likelihood approach, for MLCS2k2 and SALT2 data. 
The traditional $\chi^2$ approach gives only one value for this parameter
without
uncertainty and we represent it by the dashed vertical line in the figure. We
can see
that the discrepancy between the value obtained from the $\chi^2$ approach and
the best fit value obtained
from the likelihood approach is greater for the MLCS2k2 data. The results
are incompatible at more than 99\% confidence interval, which does not happen
for the SALT2 data. This can 
possibly be due to the fact that $\sigma_{\text{int}}$ is obtained
using only a nearby sample in the $\chi^2$ approach for MLCS2k2
while such distinction is not performed in the likelihood approach. 
This issue deserves further investigation and will be the subject of future
work.

\begin{figure*}
\begin{centering}
\includegraphics[width=0.49\textwidth]{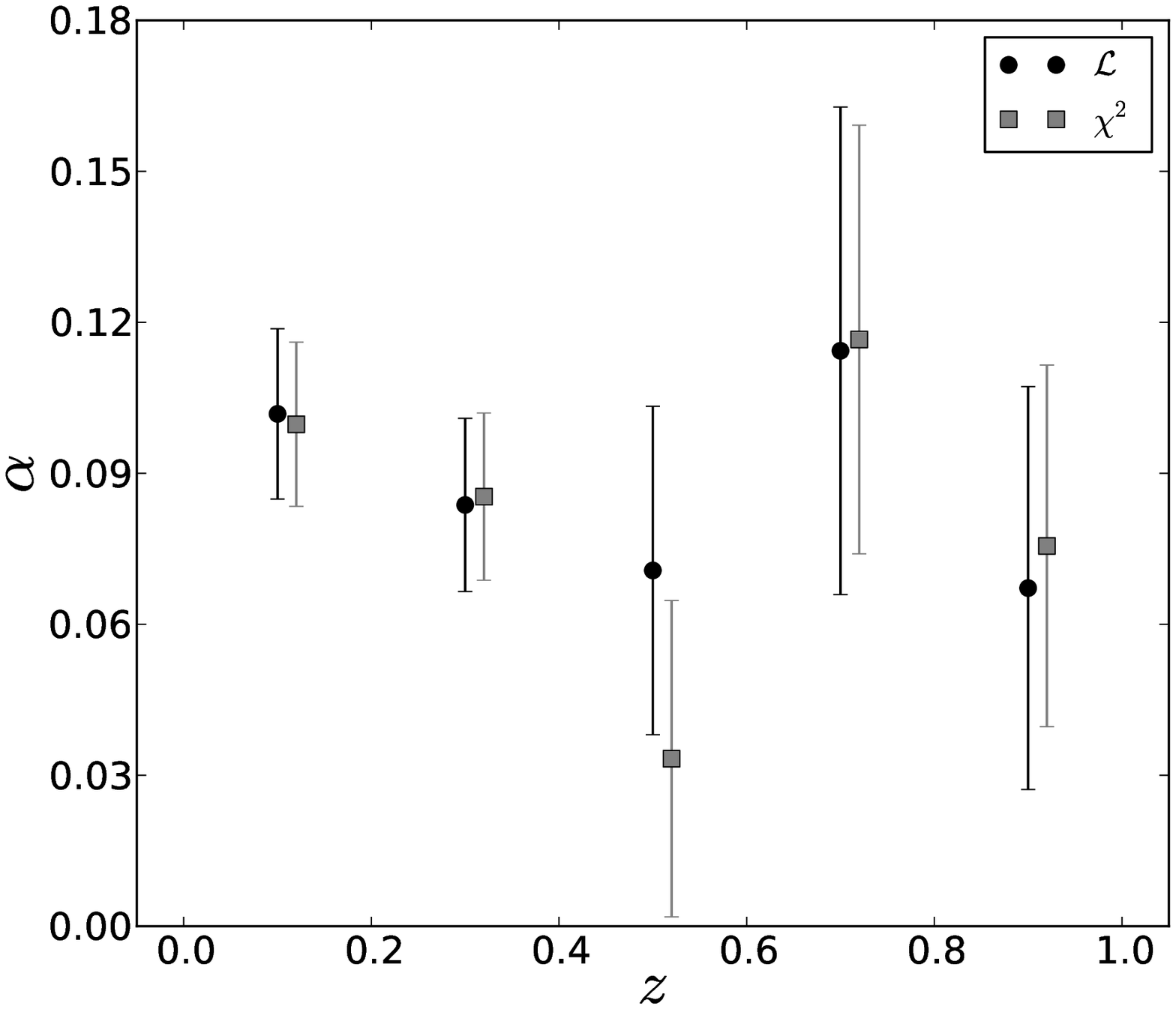}
\includegraphics[width=0.49\textwidth]{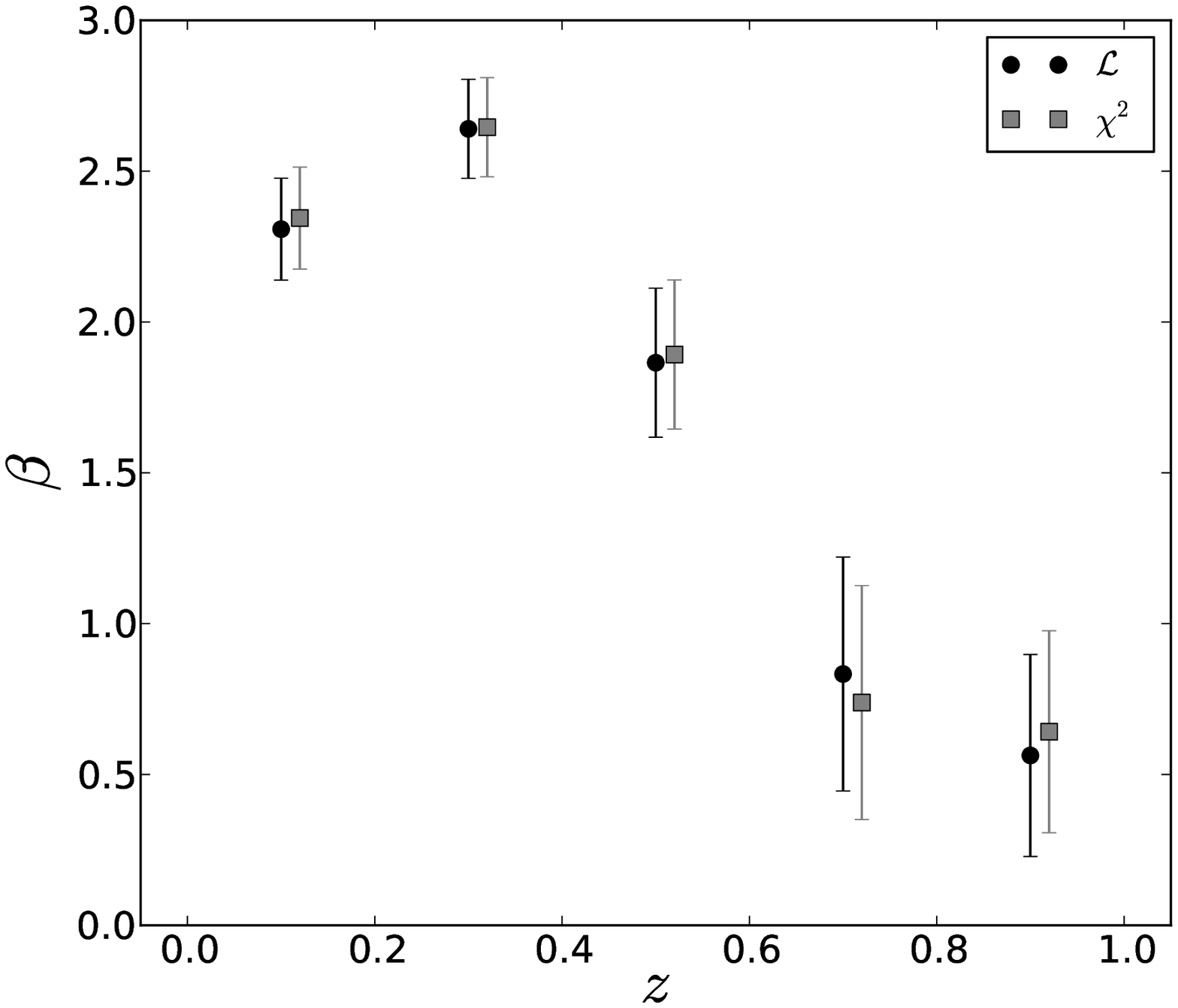}\\
\includegraphics[width=0.49\textwidth]{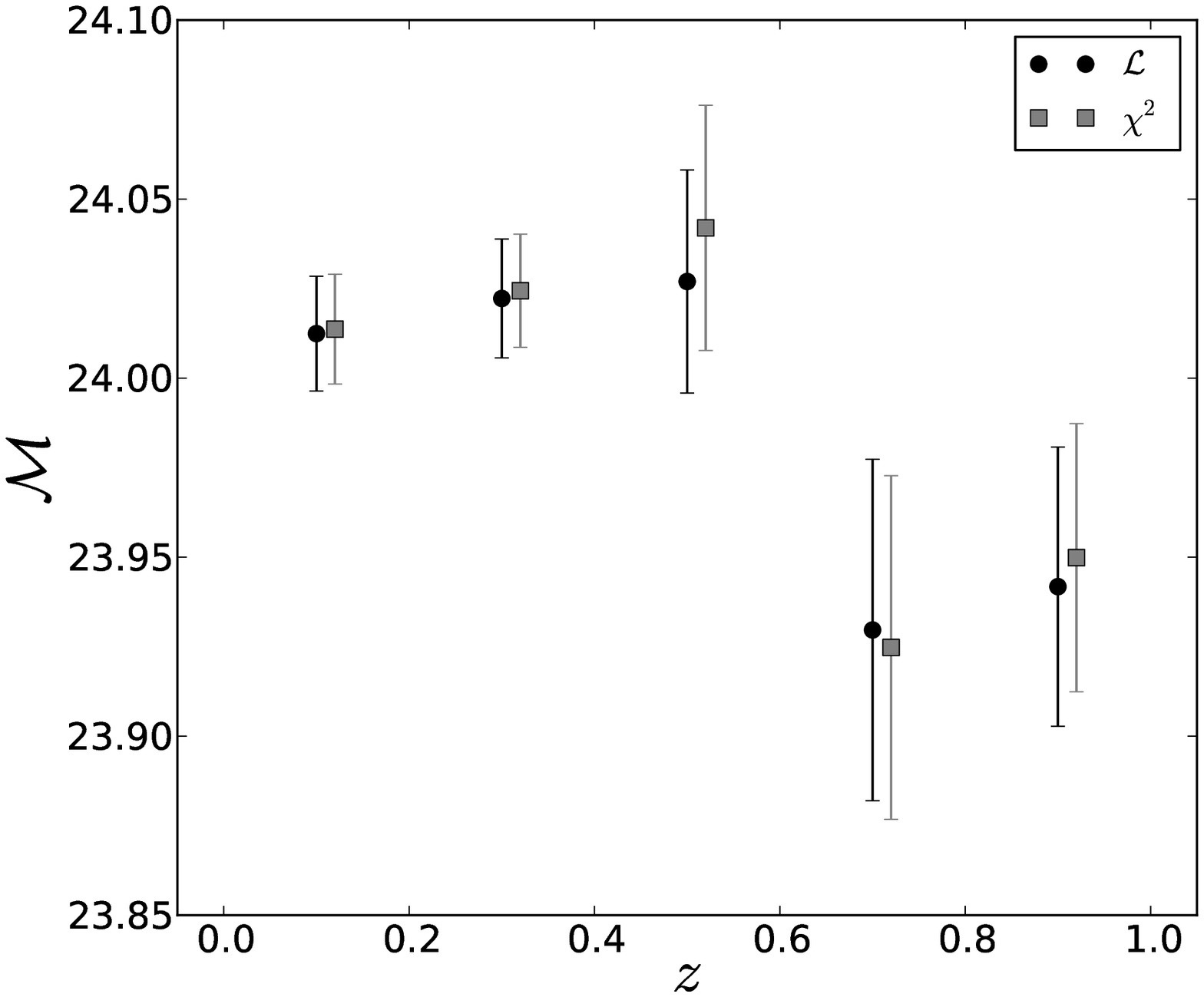}
\includegraphics[width=0.49\textwidth]{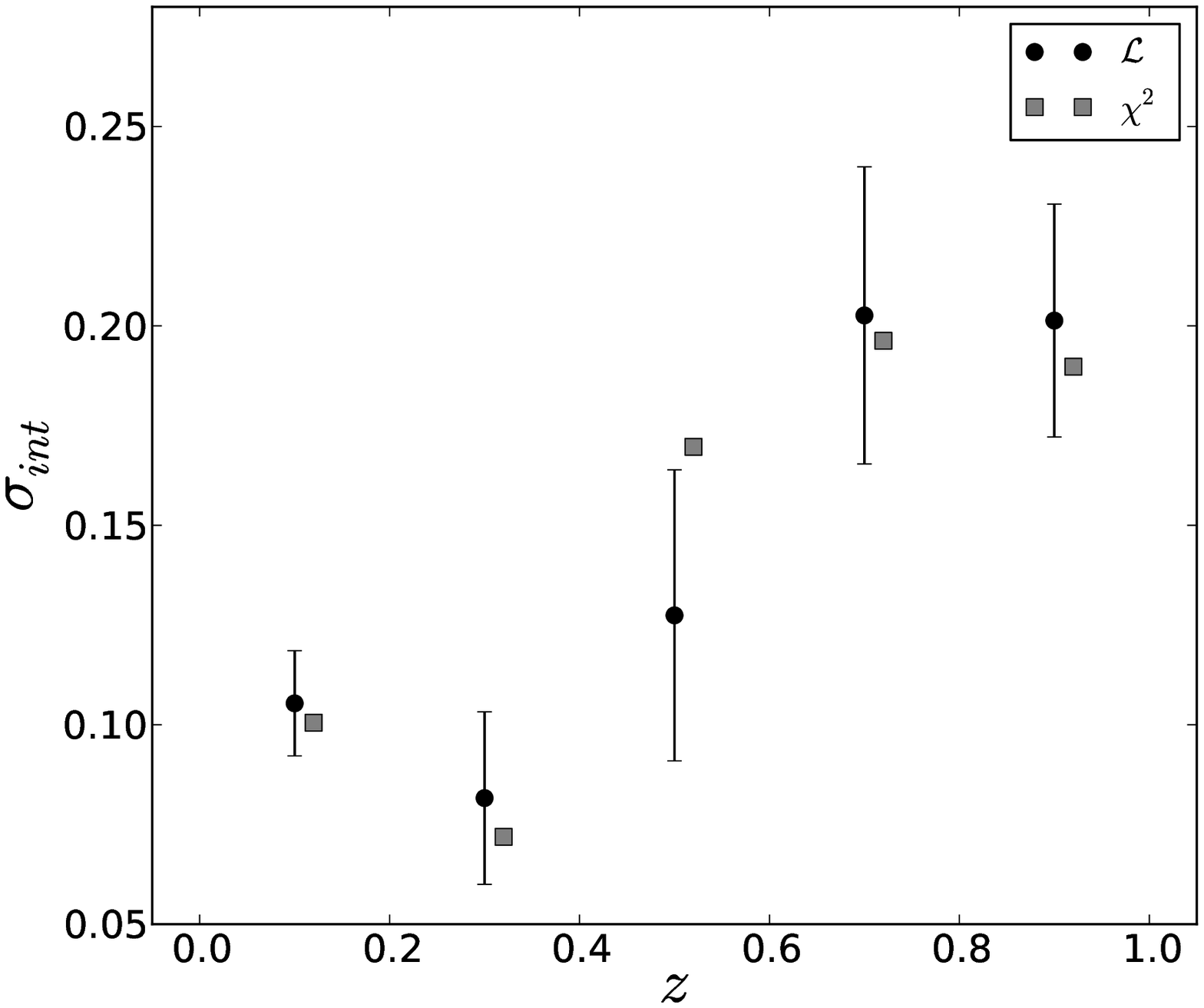}
\par\end{centering}

\caption{Evolution of SALT2 parameters $\alpha$, $\beta$, ${\cal M}$, and
$\sigma_{\rm int}$ with redshift for the SDSS compilation
for model $\Lambda$CDM. The cosmological parameters were kept
fixed in their global best fit values. 
The black circles and the gray squares are the results for the likelihood
and the $\chi^2$ approaches, respectively. The error bars represent
the 68\% confidence intervals, marginalizing over all other parameters
with flat priors.
\label{fig:SDSS_zevol}}

\end{figure*}

We also allow for a possible variation of the parameters $\alpha$, $\beta$,
$\mathcal{M}$ 
and $\sigma_\mathrm{int}$ with redshift, in the context of the SALT2 data. In
order to perform such analysis the dataset was divided in redshift bins and the
cosmological
parameters were fixed in the best-fit values obtained from the global fit, then
releasing $\alpha$,
$\beta$, $\mathcal{M}$ and $\sigma_\text{int}$ to be determined in each bin. The
results are shown in Fig.
\ref{fig:SDSS_zevol}. We found evidence of evolution for the parameter 
$\beta$, in agreement with \citep{Kessler2009}; furthermore we
also found evidence of evolution for $\sigma_{\text{int}}$, which might support
the use of a variable $\sigma_{\text{int}}$ instead of a constant one.

In addition to the SDSS compilation, we also used the Union2 compilation
\citep{Amanullah2010} and the SNLS third year data \citep{Guy2011}. However these
references
do not provide the covariances between the SALT2 parameters ($m_B^*$,$x_1$,$c$).
In order to make a fair comparison with the SDSS data, we ran again the analyses
for all three samples, this time neglecting the cross terms in Eq.
(\ref{sigsalt2}). Without taking into account the covariances, the differences
in the areas for the SDSS compilation are essentially the same. The main
difference relies on the fact that, for the F$w$CDM model, the constraints are
tighter for the $\chi^2$ approach than for the likelihood one. For the Union 2
compilation, the difference in the areas ranged from 13\% to 14\%, while
the SNLS data showed the greater discrepancies, reaching up to $\sim$53\%.

\section{Conclusion}
\label{sec:conclusions}

In this work we considered the SDSS first year compilation and compared two
different approaches 
(traditional $\chi^2$ and likelihood) to determine constraints from SN Ia
processed by two 
of the most used light curve fitters in the 
literature, the MLCS2k2 and the SALT2. The MLCS2k2 gives a 
cosmology-independent estimation
of the distance modulus for each SN Ia, with its corresponding 
variance, which can be 
directly compared with the model prediction for this quantity. 
The SALT2 is not a distance
estimator, consequently, we can only have an estimate of 
the distance modulus depending on
parameters to be obtained simultaneously with the cosmological 
ones. Furthermore, in the SNe Ia analysis, it is 
common to introduce a residual, unknown, contribution 
$\sigma_\text{int}$ to the variance 
which, in the traditional approach, is determined by imposing that the reduced 
$\chi^2$ be unity, when considering the 
full sample, in the SALT2 case, or only nearby SNe Ia, in the MLCS2k2 case.

By comparing the results obtained from the traditional $\chi^2$ approach with
the likelihood ones, 
we showed that, for current data and chosen cosmological 
models, there is a small difference in 
the best fit values and confidence contours ($\sim$ 30\% difference in area)
(cf. Figs.
\ref{fig:SDSS_LCDM_cosmopar} and \ref{fig:SDSS_wCDM_cosmopar}) in case the
MLCS2k2 fitter is adopted. 
For SALT2 the difference is less significant ($\lesssim 13\%$ difference in
areas). 
In both cases the likelihood approach gives more restrictive constraints. 
We can understand these results by observing, from Fig.
\ref{fig:SDSS_sint_dist}, that the estimated value of $\sigma_\text{int}$ in
the traditional approach, is higher than the peak of the $\sigma_\text{int}$
distribution obtained with the likelihood method. 
In fact, for MLCS2k2 the $\sigma_\text{int}$ value obtained with the traditional
$\chi^2$ approach is outside the 99\% confidence level of the distribution
obtained with the likelihood method. For SALT2 it is of the order of 68\%
confidence level and this might explain why for SALT2 the differences between
these two approaches are less significant. We should also remark that we noticed
that the covariance between the SALT2 parameters ($m^*_B,x_1,c$) has an
important role in the above result since we obtained, for the 68\% confidence
level contour, an area ratio of 1.13, when considering the covariance, and of
0.95, when neglecting the covariance.

We also studied the possible evolution of the SALT2 parameters $\alpha$, 
$\beta$ and $\mathcal{M}$
with the redshift, splitting the samples in redshift bins 
and performing the fit separately
for each one. In this situation, we found evidence of 
evolution in the parameter $\beta$ and also in $\sigma_{\text{int}}$. 

While this paper was in preparation two articles addressing 
the issue of using $\chi^2$
with unknown variances appeared in the arXiv. The first one \citep{Kim2011}
focused 
on the question of the determination of $\sigma_\text{int}$, using simulated
data for the distance modulus. The second one \citep{March2011} studied the SALT2
case and proposed a sophisticated Bayesian analysis. We did not find in our work
with the SDSS first year compilation any kind of catastrophic biases of
parameters when adopting the likelihood approach, a possibility suggested in
\citep{March2011}.

In summary, we used current data to compare the traditional $\chi^2$ and the
likelihood approaches 
to determine best fit and confidence regions from SN Ia. We argued that when 
the variance is not 
completely known, minimizing the traditional $\chi^2$ is not formally equivalent
to 
maximizing the likelihood function, since the normalization of the likelihood, 
assumed Gaussian, is also a function
of parameters to be determined. We conclude suggesting the adoption of the
likelihood framework instead of the traditional $\chi^2$ one, since it
is more straightforward, numerically more efficient and self-consistent.

\section{Acknowledgments}

We would like to thank Masao Sako for helpful discussions and suggestions. B. L.
L. and I. W. also thank CNPq, Brazil, and S. E. J. also thanks ICTP,
Italy, for support.

\end{document}